\documentclass{article}
\usepackage{latexsym,amsfonts}
\title{On algebro-geometric Poisson brackets for
the Volterra lattice}
\author{A.~P.~Veselov%
\thanks{Department of Mathematical Sciences,
Loughborough University, Loughborough, LE11 3TU, UK.
{\tt e-mail: A.P.Veselov@lboro.ac.uk}}, 
A.~V.~Pensko\"\i\thanks{Faculty of Mechanics and Mathematics, Department
of Higher Geometry and Topology, Moscow State University, Vorob'ievy Gory,
Moscow, 119899, Russia. {\tt e-mail: penskoi@mech.math.msu.su}}}
\date{}
\begin{document}
\maketitle

\abstract{
A generalization of the theory of algebro-geometric
Poisson brackets on the space of finite-gap
Schr\"odinger operators, developped
by S. P. Novikov and A. P. Veselov, to the case of periodic zero-diagonal
difference operators of second order is proposed.
A necessary and sufficient condition for such a bracket to be
compatible with higher Volterra flows is found.}

\smallskip

\noindent\textbf{AMS MSC 34G20, 34L40}

\bigskip

In the theory of integrable systems there exists a remarkable phenomenon:
variables, natural from the point of view of spectral theory and the
algebraic geometry, have ``nice'' symplectic properties.
H.~Flaschka and D.~McLaughlin~\cite{flashka}
seem to be the first to recognize this
in the examples of the KdV equation and the Toda lattice.
An attempt to formulate this phenomenon in a mathematically correct form
led S.~P.~Novikov and one of the authors (see~\cite{vn},\cite{SP})
to the notion of algebro-geometric Poisson bracket on the total
space of the bundle of hyperelliptic curves  (or the space of finite-gap
Schr\"odinger potentials).

The aim of this work is to generalize the result of~\cite{vn}
to the case of periodic difference second order operators of the form
\begin{equation}
(L\psi)_k=a_{k+1}\psi_{k+1}+a_k\psi_{k-1}.
\label{operator}
\end{equation}
These operators are closely connected with the theory of the Volterra
lattice
\begin{equation}
\dot{c}_k=c_k(c_{k+1}-c_{k-1}),\quad c_k=a_k^2,
\label{volterra}
\end{equation}
also known as ``the discrete KdV equation''~(see~\cite{Man},~\cite{soliton}).

Interest in such a generalization was brought about, in particular,
by the fact that the corresponding spectral curves have an additional
symmetry.
As a result, the quantity of the poles of the eigenfunction (a discrete version
of the Baker-Akhiezer function) is twice as the quantity
of angle variables.
Therefore, the general recipe, proposed for the fist time apparently by
E.~K.~Sklyanin~\cite{sklyanin}, which offers the coordinates of these poles
as coordinates of separation of variables, does not
apply (at least literally).

In our case one can choose (in many ways) exactly half of the poles;
in these coordinates the canonical 1-form has the ``separated'' form.

Note that in analogous though much more complicated
case of the Kowalewski top~\cite{brstn}, where the spectral curve
also has a symmetry, the similar question is not answered yet
as far as the authors know.

Recall primary facts on spectral properties of periodic
difference operators~(\ref{operator}) (see~\cite{DKN},~\cite{veselov})
with, in general, complex coefficients:
$$
\begin{array}{c}
a_{n+1}\psi_{n+1}+a_n\psi_{n-1}=\lambda\psi_n, \\
a_{n+T}=a_n.
\end{array}
$$
We confine ourselves to the case of an odd period
$T,$ $T=2N+1.$
In an even case the spectral curves have other geometry of symmetries;
this fact leads to significant differences and additional
difficulties in the appropriate theory (see~\cite{veselov}).

Consider Bloch eigenfunctions
$\psi$ such that $\psi_{n+T}=\rho\psi_n$ and $\psi_0=1.$
Floquet multiplier $\rho$ is determined by the equation
\begin{equation}
\rho+\frac{1}{\rho}=\Delta(\lambda),
\label{floquet}
\end{equation}
where $\Delta(\lambda)$ is the trace of the monodromy matrix (see~\cite{DKN}),
$\Delta(\lambda)$ is a polynomial of degree $2N+1:$
\begin{equation}
\Delta(\lambda)=\sum_{i=0}^{N}(-1)^iI_i\lambda^{2N+1-2i},
\label{delta}
\end{equation}
where $I_i$ is defined as follows.

Let $\hat{T}=\{0,\dots,T-1\}.$
A subset $I\subset\hat{T}$ is called totally disconnected if
$\forall i_1, i_2\in I\quad i_1-i_2\ne 1\,(\mbox{mod}\,T).$
Then $I_0=(\prod\limits_{i=0}^{T-1}a_i)^{-1},$
$$
I_i=(\prod\limits_{i=0}^{T-1}a_i)^{-1}\times\sum_{\lefteqn{\begin{array}{c}
\scriptstyle |I|=i, I\,\mbox{\scriptsize is totally disconnected}, \\
\scriptstyle I=(j_1,\dots,j_i)
\end{array}}}a^2_{j_1}\dots a^2_{j_i}
\quad \mbox{where}\,\,i=1,\dots,N.
$$

In particular, $I_N$ has the following form:
$$
I_N = I_0 \sum_{k=0}^{2N} a_k^2 a_{k+2}^2 a_{k+4}^2 ...a_{k+2N-2}^2
$$
(where all indices are mod $2N+1$).

Consider a spectral curve $\Gamma:$
$$
y^2=\left(\frac{\Delta(\lambda)}{2}+1\right)%
\left(\frac{\Delta(\lambda)}{2}-1\right),%
\quad y=\rho-\frac{\Delta(\lambda)}{2}.
$$
If $\Gamma$ is nonsingular then Bloch function $\psi_n$
is a meromorphic function on $\Gamma.$
A set of such operators that $\Gamma$ is nonsingular
is a domain $U,$ open in Zarisski topology.

A divisor $\cal D$ of poles of $\psi_n$
does not depend on $n$ and it is invariant under
an involution $\sigma: \Gamma\rightarrow\Gamma,$
$\sigma(y,\lambda)=(-y,-\lambda).$
At ``infinities'' $P_+$ and  $P_-:$
$y\approx\pm\frac{\Delta(\lambda)}{2}, \lambda\rightarrow\infty,$
the function $\psi_n$ has a pole and a zero of order
$n$ respectively. These properties define
the function $\psi_n$ by a given divisor $\cal D$ and curve $\Gamma$
uniquely (see explicit formulae in terms of $\Theta$-functions
of genus $N$ in~\cite{veselov}).

Note that the coordinates $\lambda_i$ of the poles of $\psi_n$
have a natural spectral sense: they are eigenvalues
of a spectral problem with zero boundary conditions:
$\psi_0=\psi_T=0.$

Consider a variety $B^{N+1}\subset{\Bbb C}^{N+1},$
consisting of such $I_0,\dots,I_N,$ that the corresponding
curve $\Gamma$ is nonsingular, i.e. such that
the polynomial $\frac{\Delta^2(\lambda)}{4}-1$ has no multiple root.
There exists a natural bundle
$E^{2N+1}\!\!\!\!\stackrel{F}{\longrightarrow} B^{N+1}$
where the fiber  $F\subset S^{2N}\Gamma$ is a space of divisors
$\cal D$ on $\Gamma$ consisting of $2N$ points
such that $\sigma({\cal D})={\cal D}.$
Then $E^{2N+1}$ coincides with $U,$ i.e.
$E^{2N+1}$ is a space of periodic operators of the form~(\ref{operator})
of period $T=2N+1,$ such that the corresponding spectral curve
is nonsingular.

There are two remarkable compatible Poisson brackets
on the space of operators~(\ref{operator}):
quadratic
\begin{equation}
\{c_i,c_j\}_1=c_ic_j(\delta_{i+1,j}-\delta_{j+1,i})
\label{1}
\end{equation}
and cubic
\begin{equation}
\{c_i,c_j\}_2=c_ic_j(c_i+c_j)(\delta_{i+1,j}-\delta_{j+1,i})%
+c_ic_{i+1}c_{i+2}\delta_{i+2,j}-c_{i}c_{i-1}c_{i-2}\delta_{i-2,j}
\label{2}
\end{equation}
(see~\cite{ft}).

Both brackets are degenerate; it is easy to prove that
the function $I_0$ is in involution with any function on the space of operators
of the form~(\ref{operator}) with respect to the first Poisson bracket
(i.e. $I_0$ belongs to an annulator of the first bracket).
An anulator of the second bracket is generated by the function
$I_N$ (see~\cite{ya}).

It was proved in \cite{ya} that coordinates of the poles
$\lambda_1,\dots,\lambda_N,$ $-\lambda_1,\dots,-\lambda_N$
are in involution with respect to both brackets.
Also in~\cite{ya} were found
variables canonically conjugate to
$q_1=\lambda_1,\dots,q_N=\lambda_N.$
In the case of the first bracket it is
$p_k=\frac{2\ln\rho_k}{\lambda_k},$ where
$\rho_k=\rho(\lambda_k)$ is the corresponding Floquet
multiplier~(\ref{floquet}).

This result can be treated as a corresponding  analogue of
Flaschka-McLaughlin theorem~\cite{flashka}.

Note that a choice of
$\lambda_1,\dots,\lambda_N$
is not unique;
one can always change
$\lambda_k$
to
$-\lambda_k.$
Under this transformation $p_k$ also changes:
$$
p_k\rightarrow -p_k+\frac{C}{\lambda_k},
$$
this transformation does not change the commutative relations.

In the case of the second bracket the same result is true for
$p_k=\frac{2\ln\rho_k}{\lambda_k^3}$ (see~\cite{ya}).

Analogously to the paper~\cite{vn} let us introduce
algebro-geometric Poisson brackets on the space $E^{2N+1}.$

Such a bracket is defined by a function $A(I_0,\dots,I_N),$ which is
an annulator of this Poisson bracket, and 1-form
$Q(\Gamma,\lambda)d\lambda$
on spectral curves $\Gamma$
in such a way that the canonically 1-form $pdq$ on a
symplectic sheet
$A(I_0,\dots,I_N)=const$
has the form
$$
\alpha=\sum_{i=1}^{N}Q(\Gamma,\lambda_i)d\lambda_i.
$$
Suppose the function
$Q$ is meromorphic (modulo a function of the annulator and
$\lambda$) in a neighborhood of one of ``infinities'',
for example, $P_-$ (compare with~\cite{vn}).
Also suppose
$Q$ satisfies the following property:
$Qd\lambda-\sigma^*(Qd\lambda)$
depends only on $\lambda$ and
the annulator.
This means that the corresponding 2-form $\omega=d\alpha$
on a symplectic sheet
$A=const$ is $\sigma$-invariant.

Clearly, both brackets~(\ref{1}) and~(\ref{2})
satisfy desribed properties; therefore they are
algebro-geometric.
For the first bracket
$A=I_0, Q=\frac{2\ln\rho(\lambda)}{\lambda},$
for the second bracket
$A=I_N, Q=\frac{2\ln\rho(\lambda)}{\lambda^3}.$

Consider the Volterra lattice~(\ref{volterra}):
$$
\dot{c}_i=c_i(c_{i+1}-c_{i-1}),\quad c_i=a_i^2.
$$
It is well known that
it is a hamiltonian system with respect to the bracket
$\{,\}_1$ and the hamiltonian
$H=\sum_{i=1}^{2N+1}c_i=\sum_{i=1}^{2N+1}a^2_i=\frac{1}{2}\mbox{tr}{\cal L}^2$
where
$$
{\cal L}=\left(%
\begin{array}{cccccc}
0 & a_1 & 0&\dots & \dots & a_{2N+1} \\
a_1 & 0 & a_2& \dots & \dots  & 0 \\
0 & a_2 & 0 & \dots  & \dots & 0 \\
\dots & \dots & \dots & \dots & \dots & \dots \\
0 & \dots & \dots & 0 & a_{2N-1} & 0 \\
0 & \dots & \dots & a_{2N-1} & 0 & a_{2N} \\
a_{2N+1} & \dots  & \dots & 0 & a_{2N} & 0
\end{array}%
\right)
$$
is a matrix in well-known Lax representation for the Volterra lattice.
The corresponding integrals
$J_0=\sum_{i=1}^{2N+1}\ln a_i=\frac{1}{2}\sum_{i=1}^{2N+1}\ln c_i,$
$J_k=\frac{\mbox{tr}{\cal L}^{2k}}{2k}, k=1,\dots,N$ are in involution
with respect to both Poisson brackets~(\ref{1}) and~(\ref{2}).
For example, it follows from the Lenard-Magri scheme.
Indeed,
for an arbitrary function $f$ relations
$$
\begin{array}{l}
\{I_0,f\}_1=0,\\
\{I_k,f\}_2=-\{I_{k+1},f\}_1,\quad k=0,\dots,N-1,\\
\{I_{N},f\}_2=0,
\end{array}
$$
follow from the theorem~3 of the paper~\cite{ya}.
We can rewrite these relations in terms of the generating
function $\Delta(\lambda):$
$$\lambda^2\{\Delta(\lambda),f\}_2=%
\{\Delta(\lambda),f\}_1.$$
It follows that
$$\lambda^2\{\ln\Delta(\lambda),f\}_2=%
\{\ln\Delta(\lambda),f\}_1.$$
It is sufficient to apply the next claim to
complete the proof.

\noindent{\sc Claim.}
There exists following expansion:
\begin{equation}
\ln\Delta(\lambda)=(2N+1)\ln\lambda-J_0-\sum_{k=1}^N\lambda^{-2k}J_k%
\quad(\mbox{mod}\,\lambda^{-2N-1}).
\label{razl}
\end{equation}

\noindent{\sc Proof.}
Let $\lambda_i, i=1,\dots,2N+1,$ be eigenvalues of the operator $\cal L.$
It is clear that they are roots of the equation
$\Delta(\lambda)=2.$
Let
$$
s_k=\sum_i\lambda_i^k,\quad
\sigma_k=\sum_{i_1<i_2<\dots<i_k}\lambda_{i_1}\dots\lambda_{i_k},
$$
i.e. $\{s_k\}$ and $\{\sigma_k\}$ are standard bases in the space
of symmetrical polynomials in
$\lambda_1,\dots,\lambda_{2N+1}.$
Then
$$
J_k=\frac{1}{2k}s_{2k},\quad\sigma_{2k}=(-1)^k\frac{I_k}{I_0},%
\quad\sigma_{2k-1}=0,\quad k=1,\dots,N.
$$
With the help of the standard formula connecting $s_k$ and $\sigma_k$
we obtain
$$
J_k=\sum(-1)^{j_2+j_4+\dots}%
\frac{(j_1+\dots+j_N-1)!}{j_1!\dots j_N!}%
\left(\frac{I_1}{I_0}\right)^{j_1}\!\!\!\!\dots%
\left(\frac{I_N}{I_0}\right)^{j_N}\!\!\!\!,\quad k=1,\dots,N,
$$
where we sum over all non-negative entire numbers
$j_1,\dots,j_N,$ such that $j_1+2j_2+\dots+Nj_N=k.$
On the other hand
$$
\ln\Delta(\lambda)=\ln I_0+(2N+1)\ln\lambda+%
\ln(1+\sum_{i=1}^N(-1)^i\frac{I_i}{I_0}\lambda^{-2i}).
$$
It is sufficient to use the expansion of logarithm and
the formula of $J_k$ to complete the proof. $\Box$

Let us define the higher Volterra flows by the formulae
$$
\dot{c}_i=\{c_i,J_k\}_1=\{c_i,J_{k-1}\}_2,\quad k=1,\dots,N.
$$

It follows easily that these flows are commuting.
A distinctive property of the integrals
$J_i$ (for example, in comparison with $I_0,\dots,I_N$)
is localization of the correspondent hamiltonian flows: the right-hand side
of the equations for $\dot{c}_i$ of
$k$-th Volterra flow depends only on
$c_j$  with $j\in[i-k,i+k].$

\noindent{\sc Definition} (compare with~\cite{vn}).
An algebro-geometric Poisson bracket is called compatible with
the higher Volterra flows if all these flows are hamiltonian
with respect to this bracket.

The main result of this paper is the following theorem.

\noindent{\sc Theorem.} a) If an algebro-geometric
Poisson bracket is compatible with the higher Volterra flows
then modulo terms with coefficients depending on annulator
there exists the following expansion at
$P_-$
$$
Q(\Gamma,\lambda)=\sum_{k=1}^N2h_k\lambda^{-2k-1}%
\quad(\mbox{mod}\,\lambda^{-2N-2}),
$$
where $h_k$ is the hamiltonian of the $k$-th Volterra flow.

\noindent b) An algebro-geometric Poisson bracket is
compatible with the higher Volterra flows
if and only if derivatives of
$Q(\Gamma,\lambda)d\lambda$
along basis vector fields tangent to the level
surface of the annulator
make up a basis in the space of
$\sigma$-invariant holomorphic
differentials on $\Gamma.$

\noindent{\sc Remark.} We can replace $P_+$ by
$P_-;$ this changes signs in the expansions of $Q$.

First let us prove the following lemma.

\noindent{\sc Lemma.} There exists the following expansion at
$P_-:$
\begin{equation}
\ln\rho(\lambda)=-(2N+1)\ln\lambda+J_0+\sum_{k=1}^N\lambda^{-2k}J_k%
\quad(\mbox{mod}\,\lambda^{-2N-1}).
\label{razl2}
\end{equation}

\noindent{\sc Proof.}
$$
\lim_{\lambda\rightarrow\infty}\lambda^{2N+1}%
(\ln\rho(\lambda)+\ln\Delta(\lambda))=%
\lim_{\lambda\rightarrow\infty}\lambda^{2N+1}%
\ln\Bigl(\Delta(\lambda)\Bigl(\frac{\Delta(\lambda)}{2}-%
\sqrt{\left(\frac{\Delta(\lambda)}{2}\right)^2\!\!\!-1}\Bigr)\Bigr)=
$$
$$
=\lim_{\lambda\rightarrow\infty}\lambda^{2N+1}%
\left(\frac{\Delta(\lambda)}%
{\frac{\Delta(\lambda)}{2}+%
\sqrt{(\frac{\Delta(\lambda)}{2})^2-1}}%
-1\right)%
=\lim_{\lambda\rightarrow\infty}%
\frac{-\lambda^{2N+1}}%
{\Bigl(\frac{\Delta(\lambda)}{2}+%
\sqrt{(\frac{\Delta(\lambda)}{2})^2-1}\Bigr)^2}=0.
$$
It is sufficient to use~(\ref{razl}) to complete the proof. $\Box$

\noindent{\sc Proof of the theorem.}
Consider an algebro-geometric Poisson bracket which is compatible with the
higher Volterra flows.
 Let $t_k$ be a time corresponding to the flow
generated by $h_k,$
let $S$ be an action,
$S=\int\limits_%
{\lambda_{1,0},\dots,\lambda_{N,0}}%
^{\lambda_1,\dots,\lambda_N}%
\sum\limits_{i=1}^NQ(\lambda_i)d\lambda_i.$
It is well known that
$\frac{d}{dt_k}\frac{\partial S}{\partial h_j}=\delta_{jk},$
therefore
\begin{equation}
\sum_{i=1}^N\frac{\partial Q}{\partial h_j}(\lambda_i)%
\frac{d}{dt_k}\lambda_i=\delta_{jk}.\label{tojd}
\end{equation}
Let $A_i^j=\frac{\partial Q}{\partial h_j}(\lambda_i),%
B^i_k=\frac{d}{dt_k}\lambda_i;$ then (\ref{tojd})
is the  matrix equality
$AB=E.$ Note that $B$ does not depend on the bracket:
this is only the matrix of derivatives with respect to the flows.
It follows from this fact and the equality
$A=B^{-1}$ that $A$ does not also depend on the bracket.
Therefore for any bracket
$\frac{\partial Q}{\partial h_j}(\lambda)=%
\frac{\partial \hat{Q}}{\partial \hat{h}_j}(\lambda)$
where  $\hat{Q}$ and $\hat{h}_j=J_k$ correspond to the bracket~(\ref{1}).
It follows from~(\ref{razl2}) that
$\hat{Q}=\sum_{i=1}^N2\hat{h}_i\lambda^{-2i-1}%
\,\,(\mbox{mod}\,\lambda^{-2N-2}).$
Therefore
$\frac{\partial Qd\lambda}{\partial h_j}(\lambda)=%
\frac{\partial \hat{Q}d\lambda}{\partial \hat{h}_j}(\lambda)=%
2\lambda^{-2j-1}d\lambda\,\,(\mbox{mod}\lambda^{-2N-2}),$
this leads to a).

It is easy to prove that the derivatives
$\frac{\partial\hat{Q}}{\partial I_j}|_{I_0=const}=%
\frac{\lambda^{2N-2j}d\lambda}{y}$ form a basis
in the space of $\sigma$-invariant holomorphic differentials on $\Gamma.$

To complete the proof of b) we must prove that
if the derivatives of
$Q(\Gamma,\lambda)d\lambda$ along
basis vector fields tangent to the level
surface of the annulator
make up a basis in the space of
$\sigma$-invariant holomorphic
differentials on $\Gamma$ then
the correspondent algebro-geometric Poisson bracket
is compatible with the higher Volterra flows.
Consider the expansion of
$Q(\Gamma,\lambda)d\lambda$ at $P_-:$
$$
Q(\Gamma,\lambda)=\sum_{k=-M}^{\infty}2\xi_k(\Gamma)\lambda^{-(2k+1)}
$$
(modulo annulator).
The derivatives of
$Q(\Gamma,\lambda)d\lambda$ are holomorphic,
therefore, all $\xi_k(\Gamma)$  with $k\le0$ are in the annulator
and the functions
$\xi_k(\Gamma)=\xi_k(I_0,\dots,I_N), k=1,\dots,N,$
where  $A(I_0,\dots,I_N)$ is fixed, are functionally independent because
these derivatives make up the basis in the space of holomorphic
$\sigma$-invariant forms.
Let us choose them as hamiltonians
$h_k=\xi_k(\Gamma),$ therefore using~(\ref{tojd})
we obtain as the correspondent hamiltonian flows the higher Volterra flows.
This completes the proof.
$\Box$

Let us illustrate this theorem on the example
of the cubic bracket.
In this case $A=I_N,$ $Q=\frac{2\ln\rho(\lambda)}{\lambda^3}.$
Consider the expansion at $P_-:$
$$
Q(\Gamma,\lambda)=\frac{2\ln\rho(\lambda)}{\lambda^3}=%
\frac{2}{\lambda^3}\left(-(2N+1)\ln\lambda+%
J_0+\sum_{k=1}^{N}\lambda^{-2k}J_k\right)+\dots
$$

The function $-\frac{2}{\lambda^3}(2N+1)\ln\lambda$ belongs to
annulator, therefore
$$
Q(\Gamma,\lambda)=%
\sum_{k=0}^{N-1}\lambda^{-2k-3}2J_k \,\,(\mbox{mod} \lambda^{-2N-2}) =%
\sum_{k=1}^{N}\lambda^{-2k-1}2J_{k-1} \,\,(\mbox{mod} \lambda^{-2N-2})
$$
modulo terms with coefficients from annulator.
But $J_{k-1}$ is the hamiltonian $h_k$ of the $k$-th Volterra flow
with respect to the cubic bracket, i.e. we obtain explicitly
the a) statement of the theorem.

Symplectic sheets are determined by the condition $I_N=const.$
As basis tangent vector fields we can use
$\frac{\partial}{\partial I_k}, k=0\dots,N-1.$
Therefore
$$
\frac{\partial Q}{\partial I_k}\,d\lambda|_{I_N=const}=%
\frac{\partial}{\partial I_k}%
\frac{2\ln\rho(\lambda)}{\lambda^3}\,d\lambda|_{I_N=const}=
$$
$$
=\frac{\partial}{\partial I_k}%
\frac{2}{\lambda^3}\ln\left(\frac{\Delta(\lambda)}{2}\pm%
\sqrt{\left(\frac{\Delta(\lambda)}{2}\right)^2-1}\right)%
\,d\lambda|_{I_N=const}=
$$
$$
=\frac{2}{\lambda^3}%
\frac{1}{\pm\sqrt{\left(\frac{\Delta(\lambda)}{2}\right)^2-1}}%
\frac{1}{2}%
\frac{\partial\Delta(\lambda)}{\partial I_k}%
\,d\lambda|_{I_N=const}=
$$
$$
=\frac{1}{\lambda^3}%
\frac{1}{\pm\sqrt{\left(\frac{\Delta(\lambda)}{2}\right)^2-1}}%
(-1)^k\lambda^{2N+1-2k}\,d\lambda=%
=(-1)^k\frac{\lambda^{2(N-k)-2}}{y}d\lambda,
$$
where the sign plus or minus depends on the sheet, $k=0,\dots,N-1.$
Clearly, we obtain the standard basis in the space of holomorphic
$\sigma$-invariant differentials,
i.e. we obtain explicitly
the b) statement of the theorem.

\subsection*{Addendum\footnote{added to the electronic version 24 november 2000.}}

We would like to mention that nowadays there exists a universal approach
to the Hamiltonian theory of integrable systems based on Lax representations.
It has been developed by I.M.Krichever and D.H.Phong in the fundamental papers
\cite{KP1, KP2}. It would be interesting to investigate what it gives for the
Volterra system.

\end{document}